# Regional and Global Science: Publications from Latin America and the Caribbean in the *SciELO Citation Index* and the *Web of Science*

# Ciencia Regional y Global: Publicaciones de América Latina y el Caribe en el *SciELO Citation Index* y el *Web of Science*

Gabriel Velez-Cuartas*,[1], Diana Lucio-Arias[2,] and Loet Leydesdorff [3]

[1] Grupo de Investigación Redes y Actores Sociales, Departamento de Sociología, Facultad de Ciencias Sociales y Humanas, Universidad de Antioquia; Calle 70 No. 52-21, Medellín, Colombia; gabrielvelezcuartas@gmail.com;
*Corresponding author

[2] Colombian Observatory of Science and Technology, Bogota, Colombia; dlucioarias@gmail.com

[3] University of Amsterdam, Amsterdam School of Communication Research, PO Box 15793, 1001 NG Amsterdam, the Netherlands; loet@leydesdorff.net

**Abstract**

We compare the visibility of Latin American and Caribbean (LAC) publications in the Core Collection indexes of the *Web of Science* (WoS) —*Science Citation Index Expanded*, *Social Sciences Citation Index*, and *Arts & Humanities Citation Index*—and the *SciELO Citation Index* (SciELO CI) which was integrated into the larger WoS platform in 2014. The purpose of this comparison is to contribute to our understanding of the communication of scientific knowledge produced in Latin America and the Caribbean, and to provide some reflections on the potential benefits of the articulation of regional indexing exercises into WoS for a better understanding of geographic and disciplinary contributions. How is the regional level of SciELO CI related to the global range of WoS? In WoS, LAC authors are integrated at the global level in international networks, while SciELO has provided a platform for interactions among LAC researchers. The articulation of SciELO into WoS may improve the international visibility of the regional journals, but at the cost of independent journal inclusion criteria.

**Resumen**

Comparamos la visibilidad de las publicaciones de América Latina y el Caribe (LAC) en la colección principal de índices de *Web of Science (WoS) –Science Citation Index Expanded, Social Science Citation Index, y Arts & Humanities Citation Index (SciELO CI)*, el cual fue integrado en la plataforma de *Web of Science* en 2014. El propósito de esta comparación es contribuir al entendimiento de la comunicación del conocimiento científico producido en Latinoamérica y el Caribe, y presentar algunas reflexiones sobre el potencial beneficio de la articulación entre los ejercicios regionales de indexación regional con *Web of Science* para un mejor entendimiento de las contribuciones geográficas y disciplinarias. ¿Cómo está el nivel regional de *SciELO CI* comparado con el global de *WoS*? En *WoS*, los autores de Latinoamérica y el Caribe están integrados en el nivel global de las redes internacionales; en *SciELO CI*, ha proveído una plataforma de interacción entre investigadores de América Latina y el Caribe. La articulación de *SciELO* en el *Web of Science* podría mejorar la estandarización internacional (por ejemplo, de referenciación) en las revistas regionales, pero al precio de perder independencia en los criterios de inclusión de las propias revistas.

**Palabras clave:** revistas, bases de datos, índice, *SciELO, WoS*, América Latina, Caribe

**Introduction**

The development of scientific capacities in Latin America and the Caribbean faces multiple challenges, including limited investments in R&D, low participation of researchers and qualified personnel in the labor force, insufficient infrastructure and specialized laboratories, and inadequate circulation and visibility of research results. *Nature's* (2014) special issue about research in the South American continent raised some of these issues, but also highlighted some elements that could be causing "underestimation" of LAC research. This issue was revisited again by *Nature* in 2015. In this latter article some promising fields of research were indicated, in which LAC researchers can improve their visibility in the short term.

Authors from Latin America and the Caribbean tend to publish in regional and local journals. Brazil, which accounts for half of the scientific output of the LAC region in the *Science Citation Index* (RICyT, 2013), publishes approximately 40% of its scientific production outside the Core Collection of the Web of Science (Mugnaini, Digiampetri, & Mena-Chalco, 2014). However, the inclusion of LAC-edited journals in Thomson Reuter's and Elsevier's main indexing services (WoS and Scopus) has increased over time (Testa, 2011). The number of publications with at least one author affiliated to an institution in LAC has also increased. The number of publications from all Latin American and Caribbean countries (with the exception of Venezuela) has increased during the last 15 years (Van Noorden, 2014). However, part of this growth can be explained by the increased number of regional journals included in the databases. The share of research articles from LAC countries is still approximately four percent in the indexing services, which is lower than the share of the region in the world population or world GDP. The latter has been estimated as between five and six percent (Van Noorden, 2014).

Growth in the number of LAC contributions in recognized databases of scientific publications has been interpreted as a successful integration of the region into the global system of scientific communication. This integration takes place despite a gap in the production of high-quality journals in LAC, which has been documented elsewhere (e.g. Meneghini, Mugnaini & Packer, 2006) and the predominance of Spanish and Portuguese as preferred languages by most researchers in the region. Through an Open Access platform, SciELO has provided visibility to LAC research results with important spillovers to improve the quality and reduce language barriers. First, by providing a set of clearly defined requisites to enter the platform, SciELO has disseminated international norms and quality standards among the region's editors. This has also been the case in Spain, where complyiance with SciELO's set of technical requirements and format norms requires editors to invest time in organizing their information and metadata (Fraga Medín *et al*., 2006). Second, by defining a classification and evaluation system for the journals in the region, SciELO has served as a communication system for researchers who prefer to publish in their mother tongue: in 2013, only 33.62% of the journals in SciELO had English as the main language.

SciELO's contribution to global science relies on its impact in the circulation and visibility of LAC's scientific production. Although the real impact of the SciELO exercise has yet to be measured, SciELO has become an important tool for the development of scientific capabilities in LAC during the last 15 years (Delgado-Troncoso, 2011). Its main goal has been to increase the participation of the region in "world class" scientific results, particularly through the consolidation of a regional base of high-quality scientific journals (Parker et al., 2014). The financial requirements to maintain such an updated, expanding, and relevant exercise (Aguillo, 2014), together with the potential of journals indexed in SciELO to provide a representation of LAC science, may explain the interest behind the inclusion of the regional exercise in the databases owned by Thomson Reuters (Testa, 2011).

The inclusion of SciELO in WoS has had a mixed reception in the LAC scientific community. In 2007, an alliance between Scopus and SciELO first raised expectations that all SciELO information would be included in Scopus (Elsevier, 2007). The potential impacts of the inclusion of the journals, and the ambiguity of whether all SciELO journals would be included, raised some concerns in the LAC scientific community. The negotiations thereafter about SciELO's inclusion either in Scopus or WoS were perceived by some editors of LAC journals as a "sell-out" of SciELO's principles, which generated uncertainty about the future of the regional journal structure that SciELO had aimed to consolidate.

With this paper we hope to contribute to the discussion about the role of both indexes in the LAC scientific communication. In the next section we introduce the data and methods employed in this study. The results section focuses on the differences among the indexes, specifically on the geographical and collaborative aspects, and on the disciplinary characteristics of the communications in each of them. We finish this contribution with some reflections on the challenges and opportunities of the integration of SciELO into WoS.

**Data and Methods**

Using a search query for all LAC countries AND publication year 2013 in WoS, 92,900 documents were retrieved on June 6, 2015. We did not use 2014 in order to avoid indexation delays and incomplete pictures of the yearly results. The same information was downloaded for 29,729 documents that responded to the same search query in the SciELO CI online available through WoS. The organization of this information into relational databases was possible through dedicated routines available at http://www.leydesdorff.net/scielo and http://www.leydesdorff.net/software/isi, respectively.

**Table 1\*.** Differences in the sets of LAC publications from SciELO CI and WoS Core collection

| LAC publications | SciELO CI | | | WoS Core Collection | | |
|---|---|---|---|---|---|---|
| *N* of records | 29,654 | | | 92,900 | | |
| **Statistics** | N | μ | σ | N | μ | σ |
| Authors** | 88,943 | 3.69 | 2.34 | 266,755 | 11.06 | 111.6 |
| Addresses | 10,666 | 2.33 | 1.57 | 187,036 | 3.60 | 11.68 |
| Times cited | 4,424 | 0.15 | 0.55 | 200,045 | 2.15 | 6.78 |
| Cited references | 694,935 | 28.29 | 18.8 | 2,252,759 | 36.56 | 30.1 |
| Subject Categories | 190 | 1.24 | 0.74 | 285 | 1.5 | 0.7 |
| Indexed Sources *** | 771 | 38.46 | 40.8 | 9,090 | 10.22 | 29.41 |

\* This table shows the number of authors, addresses, citations, references and subject categories listed in WoS and SciELO CI. Mean and Standard deviation derive from distribution of the articles in each one. Indexed sources are the total number of Journals. Mean and standard deviation represents the proportion of articles published in each source.

\*\* We use author, addresses and references data without normalization. Only for author forms, we assume that two author names which coincide completely in terms of the last name and at least two initial of the first name are the same form. Accent marks in author names were corrected as well.

\*\*\* We counted the number of sources containing scientific production with LAC addresses in each of the indexes; the mean and the standard deviation are based on the numbers of papers per source.

In order to assess some of the differences in the sets of data considered in this analysis, we provide some descriptive statistics in Table 1. We include the mean (μ) and the standard deviation (σ) to provide some order of magnitude and dispersion among the attributes. The differences among the types of communications included in each set are considerable. Among other things, Table 1 shows that the documents in journals indexed in WoS have on average more citations, and result more frequently from collaborations among larger numbers of authors. These are most often from European or American institutions. Furthermore, these documents are more codified (in terms of the cited references used) and, on average, have a significantly larger impact (in terms of citations received).

The mean and standard deviation of the category "sources" provides the average number of documents with LAC authors per journal or source (proceedings and books are hereby included). Although there are fewer journals in SciELO CI than in WoS (771 *vs* 9,090; see Table 1), the dispersion among the different source names is greater in SciELO CI. As expected, SciELO CI indexed journals have a larger participation of LAC authors than WoS journals: LAC authors (co-)author 75,1% of the publications in SciELO CI, while this participation is lower than 5% in WoS (in June 2015, a total of 2,352,374 documents were included in WoS with publication year 2013, and 39,477 in SciELO CI). A total of 163 of these journals are indexed in both WoS and SciELO CI.

We used the Overlaymaps Toolkit available at http://www.leydesdorff.net/overlaytoolkit (Rafols, Porter & Leydesdorff, 2010) to provide visualizations of the relations among disciplines in each of the document sets (SciELO CI and WoS Core Collection). First, we retrieve a set of documents at the WoS and SciELo CI. Then SC is assigned thorugh the function Analyze provided in the Web of Science. A list of number of articles present in each category is generated. This list can generate a map of science using Pajek in which the size of a node (SC) is proportional to the number of documents in that category (Rafols, Porter & Leydesdorff, 2010).

To reflect upon the distinctions in the collaborative nature of the communications in each index, we built co-authorship networks between countries using Pajek. Collaborations were retrieved from each pair of co-authorships presents in documents. All addresses were aggregated in five different regions and contrasted with each LAC country. We rely on these visualizations and descriptive statistics to present: (1) the dynamics of the scientific workforce (authorship, country affiliation, nature of publishing sources); (2) social integration in regional and global science (co-authorship dynamics, country and regional affiliations); and (3) intellectual organization (overlay maps) in each of the sets of documents. We expect that substantial differences between the two databases will reflect diverse goals and interests in the management of each of the indexes, as discussed above. Furthermore, these three aspects of the dynamics can explain differences between the visibility regimes of publications in both databases.

**Results**

*Authorship and country affiliation*

In this section, we provide some results about the differences between communications in the Core Collection of WoS and the recently integrated SciELO CI, focusing on the regional, collaborative and cognitive aspects underlying these communications. In Table 2, the number of records is provided in each of the sets by country of origin of the authors. In order to normalize for documents with co-authorships, we include a fractional count of the documents considering the total number of signing authors. To reflect on the position of the researcher in the list of authors, we included a column where the amount of records had an address in LAC as the affiliation of the first author.

**Table 2.** Regional distribution of papers in WoS Core collection and SciELO CI.

| Country | SciELO CI | | | WoS | | |
|---|---|---|---|---|---|---|
| | *Records* | *Fractional* | *First author* | *Records* | *Fractional* | *First author* |
| Brazil | 18,178 | 6,514.47 | 17,281 | 51,135 | 13,515.96 | 44,110 |
| Colombia | 2,801 | 1,467.52 | 2,516 | 4,996 | 1,586.22 | 3,369 |
| Chile | 2,438 | 1,315.47 | 2,154 | 8,146 | 2,628.24 | 5,402 |
| Mexico | 2,339 | 1,133.04 | 2,089 | 16,098 | 4,386.14 | 12,468 |
| Cuba | 1,852 | 947.85 | 1,666 | 1,268 | 359.5 | 870 |
| Argentina | 1,728 | 708.01 | 1,521 | 11,261 | 3,366.1 | 8,542 |
| Venezuela | 502 | 248.63 | 403 | 1,399 | 411.65 | 920 |
| Peru | 415 | 186.27 | 350 | 1,148 | 305.52 | 467 |
| Costa Rica | 387 | 200.6 | 295 | 588 | 171.05 | 267 |
| Uruguay | 92 | 43.98 | 61 | 1,005 | 278.52 | 591 |

| Ecuador       | 57 | 22.9  | 32 | 597 | 154    | 233 |
| Bolivia       | 34 | 21.57 | 21 | 101 | 14.54  | 10  |
| Guatemala     | 10 | 3.7   | 7  | 70  | 9.02   | 9   |
| Panama        | 26 | 7.34  | 15 | 439 | 120.42 | 124 |
| Puerto Rico   | 19 | 11.13 | 14 | N/A | N/A    | N/A |
| Paraguay      | 20 | 5.78  | 15 | 51  | 6.62   |     |
| El Salvador   | 10 | 4.18  | 6  | 23  | 2.94   |     |
| Nicaragua     | 15 | 7.9   | 10 | 27  | 3.93   | 5   |
| Honduras      | 3  | 0.78  | 2  | 32  | 3.67   | 5   |
| Dominican Rep.| 4  | 0.98  | 2  | 30  | 3.69   | 4   |

The divergence in the countries' participation in the scientific production of LAC can result from the degree to which a specific country has been articulated in the SciELO program and the efforts to increase the SciELO journal list of each country. The most important SciELO journal collection comes from Brazil and includes 337 journal titles; Colombia follows with a total of 184 journal titles; Mexico has 149 titles; Argentina and Chile 107 and 106 journal titles, respectively. Another explanation is the specific countries' policies and the importance attributed to national scientific journals in this context. Collazo-Reyes (2014) provides a third explanation for this divergence. He states that in the period of 2006-2009 WoS increased the number of LAC Journals included in the database from 69 to 248 titles. Latindex, which is the most comprehensive catalogue of academic journals edited in LAC, allows one to certify the differences within the region in terms of the formalization of the academic journal structure. Considering Latindex and the incremental inclusion of LAC Journals in the databases, we can observe differences in the participation of countries: Colombia has around 63% of its journals either in SciELO or Scopus, Mexico has 47%, Chile 39%, and in Argentina and Brazil just 29% of the journals listed in Latindex are in either SciELO or Scopus (Miguel, 2011).

Policy efforts to support national scientific journals vary in the region; some countries privilege international publication while others aim at balancing international visibility with support to local journals and local publishers (Vessuri *et al*., 2013). Different publication strategies are also evident from Table 2, where the effect of fractional counting seems to be more drastic for communications in journals indexed in the WoS Core Collection than in SciELO CI. In Colombia, for example, collaboration with international peers has increased the participation of authors based in the country in high-quality journals (Lucio-Arias, 2013). If we take into account the number of records, one can nonetheless argue that Colombia and Cuba envision a regional strategy due to the number of records available in SciELO CI in comparison with those available in WoS (half for Colombia and an even larger percentage for Cuba). Other countries show at least one-third or more entries in SciELO CI compared with those in WoS.

With respect to the first-author column in Table 2, it is remarkable that more than 2/3 of the papers have a LAC researcher as first author. The participation of LAC researchers as first authors in the global production seems to be due to former post-doc and PhD students working in large international groups, and to the collaboration between research institutes in LAC and North American and European programs. LAC researchers participate in global research by participating in large research programs.

*Nature of publishing sources*

We expect some of the differences in the communications to result from differences in the journals included in each of the indexes. Open access journals, which are supported by SciELO, result from the lack of with academic interest on the part of commercial publishers in the LAC region (Vessuri et al., 2013). To explore this issue further, we derive Table 3 from the publisher's data available in both WoS Core Collection and SciELO CI. The classification is based on a search strategy for semantic elements that could distinguish companies (Ltd., Pub., Press, Edit, Verlag, Inc.), popular commercial publishers (Springer, Elsevier, Wiley,

Taylor & Francis), and academic sources (Univ, Asso, Inst). This search strategy allowed us to classify almost all the publishing sources available in the databases and compare them in terms of overall frequencies and participation.

**Table 3**. Nature of publishing sources.

| Semantic root* | WoS | % | SciELO | % |
|---|---|---|---|---|
| Ltd | 1.307 | 16,61 | 2 | 0,25 |
| Pub | 905 | 11,50 | 4 | 0,50 |
| Press | 640 | 8,13 | 1 | 0,13 |
| Edit | 93 | 1,18 | 34 | 4,28 |
| Verlag | 182 | 2,31 | 0 | 0,00 |
| Inc | 1.027 | 13,05 | 0 | 0,00 |
| Springer | 941 | 11,96 | 0 | 0,00 |
| Elsevier | 1.299 | 16,50 | 1 | 0,13 |
| Wiley | 840 | 10,67 | 0 | 0,00 |
| Taylor & Francis | 406 | 5,16 | 0 | 0,00 |
| Univ | 278 | 3,53 | 381 | 47,92 |
| Asso, Soc | 793 | 10,07 | 222 | 27,92 |
| Inst | 77 | 0,98 | 106 | 13,33 |
| Total journals | 7.871 | | 795 | |

* Although the semantic roots could overlap (for example, "Wiley-Blackwell Inc." or "Springer Verlag"), we assigned only one of the semantic roots to each journal)

According to Table 3, most publishing sources in WoS with documents from LAC authors come from commercial publishing companies. While the four largest companies publish almost 50% of the WoS journals with contributions from LAC authors, publication media issued by universities and professional associations roughly explain 13,6% of these journals. It is worth mentioning that in the case of WoS, journals from professional associations are often published by commercial publishing houses, for example Wiley for the case of JASIST, and therefore are considered in Table 1 as commercial publishing rather than professional. This is opposite to what we find in SciELO, where journals from universities, institutions, and associations contain the majority of contributions by LAC authors (89.1%).

This difference in the nature of publishing houses can have important effects on the nature of the scientific communications in each set of documents. Commercial publishing companies may be more willing to invest in communication strategies to increase visibility and prestige and improve indexation probabilities and positions. For academic institutions, similar strategies based on public relations and communication may well be less common. In this sense, the inclusion of SciELO CI into WoS appears as a win-win strategy: SciELO-indexed journals gain in visibility, while WoS gains in regional coverage.

*Co-authorship dynamics and affiliations*

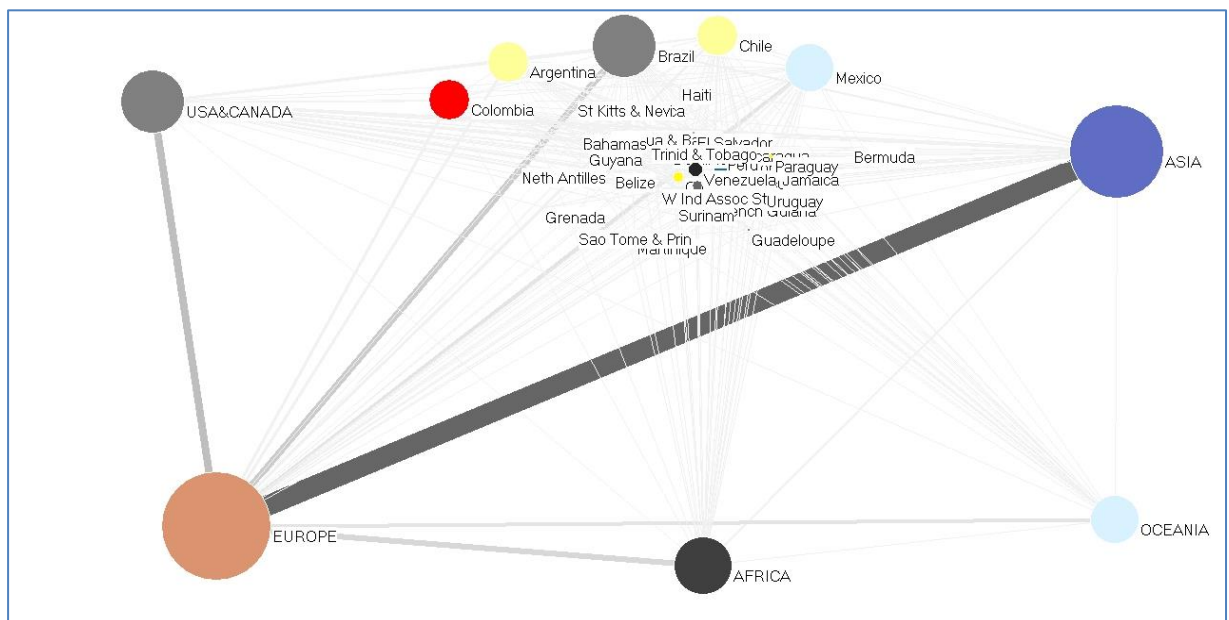

**Figure 1.** International Collaboration including LAC authors in the WoS Core Collection.

The alliances and collaborations are reflected in important differences in the networks of collaboration that emerge from scientific communications with at least one author from LAC in each of the two indexes (see Figures 1 and 2; Tables 4 and 5). The co-authorship map based on WoS data (Figure 1) shows a stronger integration of researchers from LAC with European and Asiatic peers than with the USA and Canada. The mediation of North American and European countries in the production of scientific knowledge in the region suggests a predominance of *global* topics of research in the WoS database. This is also suggested in Table 1, where the average number of authors in WoS announces the participation of LAC in the highly collaborative science common in research projects of considerable magnitude, like the projects in the context of CERN's accelerator. In many cases, these relations are maintained by (former) post-docs and PhD students who have spent time in these host countries.

In other words, collaboration of LAC countries with peers "from the north" dominates the scientific communications in which LAC scholars participate. Regional (LAC-LAC) collaboration seems not very relevant and even less important than collaborations with Asia, Africa, and Oceania. South-South collaboration has received a lot of attention in the political discourse (Arunchalam & Doss, 2000; Chandiwana & Ornbjerg, 2003) and has become an important issue in the development policy agenda (there is a United Nations Office for South-South cooperation with a website at http://ssc.undp.org/content/ssc.html). Nevertheless, South-South collaboration, as depicted in Figure 1, is mostly mediated by developed countries and does not necessarily represent a transfer and exchange of resources and knowledge among developing nations. Across-border collaboration among LAC countries appears weak in WoS.

**Table 4.** Collaborations in WoS documents with at least one address in LAC

| Rank | Number | Collaboration |
|------|--------|---------------|
| 1 | 83,224 | Europe-Europe |
| 2 | 52,701 | Asia-Europe |
| 3 | 51,277 | Europe-LAC |
| 4 | 20,442 | Europe-USA&Canada |
| 5 | 17,925 | USA&Canada-LAC |
| 6 | 14,986 | Europe-Africa |
| 7 | 13,268 | Asia-LAC |
| 8 | 8,926 | Europe-Oceania |
| 9 | 8,841 | Asia-Asia |
| 10 | 8,384 | LAC-LAC |
| 11 | 6,131 | Asia-USA&Canada |
| 12 | 5,465 | Asia-Africa |
| 13 | 4,155 | Africa-LAC |
| 14 | 3,181 | Oceania-LAC |
| 15 | 3,014 | Asia-Oceania |
| 16 | 1,739 | Africa-USA&Canada |
| 17 | 1,655 | Oceania-USA&Canada |
| 18 | 1,080 | USA&Canada-USA&Canada |
| 19 | 936 | Africa-Oceania |
| 20 | 840 | Africa-Africa |
| 21 | 181 | Oceania-Oceania |

Note: In Tables 4 & 5, the presence of co-authorship relations that are different from relationships between LAC countries and other regions occur due to the counting of each pair of relations that co-occur in a single document.

In Tables 4 and 5, we aggregated the LAC contribution in order to obtain a network of publications among world regions (LAC, Europe, Asia, USA+Canada, Africa, and Oceania) for the WoS Core Collection and SciELO CI, respectively. Table 4 first shows that the participation of LAC researchers in intra-European networks of collaborations is the main category in WoS. Secondly, LAC authors participate in collaboration networks with authors from Asia and Europe. Intra-LAC collaboration follows only at the 10$^{th}$ place.

In summary, international collaboration at the global level has a higher frequency than regional collaboration within LAC (Wagner, Park & Leydesdorff, 2015) on the basis of WoS data. Therefore, the role of geographical proximity in research collaboration might become more readily apparent when relying on regional indexing exercises like SciELO (cf. Ponds *et al*., 2007), because in international collaborations at the level of WoS th global network prevails.

This picture changes when considering contributions in SciELO CI indexed journals. The resulting map of collaborations (Figure 2) suggests a more pronounced strategy based on the regional conjugation of research efforts. Collaboration with Europe is mainly oriented towards Spain and Portugal, suggesting that linguistic and cultural similarities are a strong motivation to collaborate.

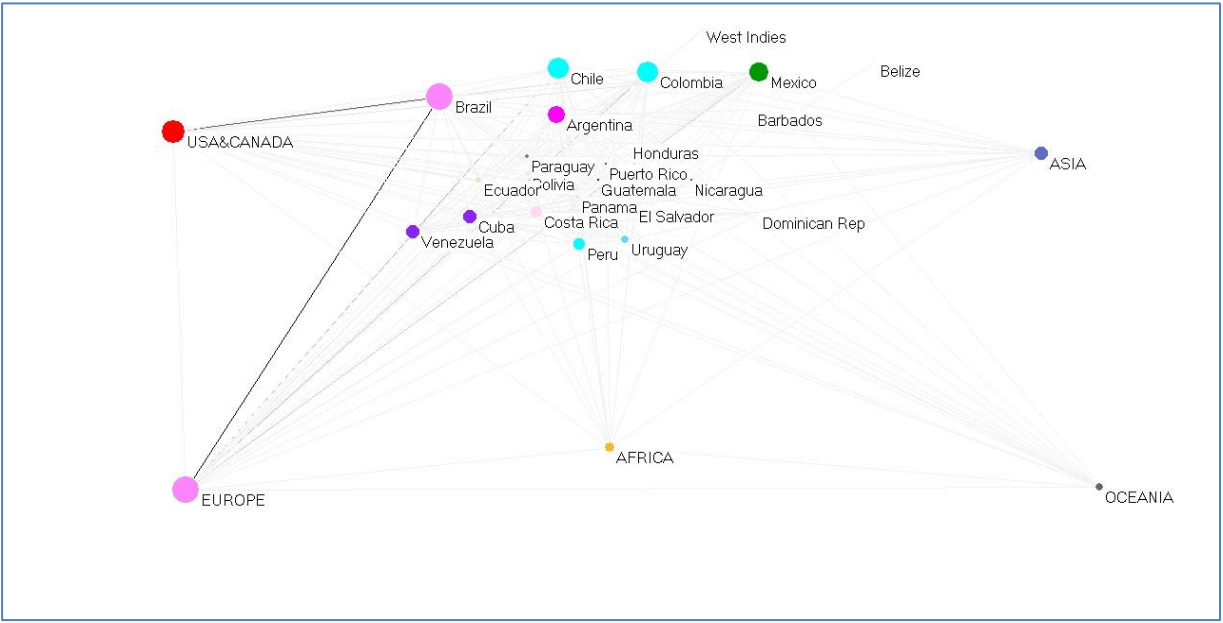

**Figure 2**. International Collaboration including LAC authors in SciELO CI.

Collaborations with Europe, and to a lesser extent with the USA and Canada, or Asia, remain strong in SciELO CI, but LAC authors seem less dependent on large-scale multinational collaborations. The more central position of LAC countries in the map suggests the importance of collaborations in strengthening and consolidating research capacities in the region. Despite the geographical proximity of the USA and Canada, Europe remains the main partner of authors in the LAC countries. Brazil, Colombia and Mexico tend to have the highest rates of collaboration with Europe and the USA. The strong collaborative ties between Mexico and the USA may reflect their geographical proximity and economic relations (NAFTA). Table 5 summarizes the results in a format directly comparable to Table 4 (that is, at the level of world regions).

**Table 5. Collaborations in SciELO documents with at least one address in LAC**

| Rank | Value | N |
|---|---|---|
| 1 | 1,300 | Europe-LAC |
| 2 | 1,217 | LAC-LAC |
| 3 | 671 | USA&Canada-LAC |
| 4 | 98 | Europe-Europe |
| 5 | 92 | Asia-LAC |
| 6 | 66 | Oceania-LAC |
| 7 | 64 | Africa-LAC |
| 8 | 56 | Europe-Asia |
| 9 | 52 | Europe-USA&Canada |
| 10 | 24 | Asia-USA&Canada |
| 11 | 18 | Africa-Europe |
| 12 | 16 | Asia-Asia |
| 13 | 9 | USA&Canada-USA&Canada |
| 14 | 7 | Oceania-Europe |
| 15 | 6 | Africa-Asia |
| 16 | 4 | Africa-USA&Canada |
| 17 | 1 | Africa-Africa |
| 18 | 1 | Africa-Oceania |
| 19 | 1 | Oceania-USA&Canada |
| 20 | 1 | Oceania-Oceania |

*It was not possible to determine the country of origin in the case of 1,161 address records.

Although they deserve further research, collaborations in SciELO seem to show more South-South cooperation than WoS-based publications. As noted, these collaborations are important as exchanges of resources and ideas within and among developing countries to solve similar development problems. In Table 5, LAC appears more visible in terms of participation in collaborations than in Table 4. Moreover, regional (LAC-LAC) collaboration is ranked in the second place after EUROPE-LAC co-authorships.

In Figure 2, collaborations within LAC, and with Africa or Asia, provide a stronger representation of South-South cooperation. We expect less mediation of the North in South-South collaborations in SciELO CI-indexed communications. However, we also find similarities with WoS-indexed contributions. The comparison between SciELO and WoS suggests that the regional strategy set by SciELO has had some impact in promoting the visibility of LAC-LAC collaboration, as they have a higher frequency in Table 5 than in Table 4.

In summary, the differences between Figures 1 and 2 suggest that communication practices differ when (a) aiming at results with international visibility than when (b) the primary goal is regional or local diffusion of scientific results through regional journals. While for WoS (Figure 1) strong ties are shown with North America and Europe, regional collaboration is more dominant in Figure 2. The major participation of the USA and Europe in Figure 1 and of Brazil in Figure 2 should be interpreted considering that these countries have the highest numbers of indexed journals in each of the respective databases.

*Overlay maps*

One advantage of the integration of the SciELO CI database into WoS is that Thomson Reuters attributes the same WoS Subject Categories (WCs) to journals in both databases. The subject categories indicate disciplinary groupings and topical sets of journals (albeit sometimes with some error; Leydesdorff & Bornmann, in press). Maps are built on the basis of a matrix of similarity measures computed from aggregated journal-journal citation relations. Rafols *et al.* (2010) provided a comprehensive map of WoS based on WCs as aggregates of journals, on which one can project any subset from WoS by using overlays to the base map.

We projected our two subsets with LAC authors on this basemap in order to visuzalize the differences in terms of cognitive categories. The map using WoS data (Figure 3) shows some dominance of the "hard" sciences, which are more likely to be published in English and in collaborations. For SciELO CI (Figure 4) the disciplinary participation seems to favor the social, health, and agricultural sciences.

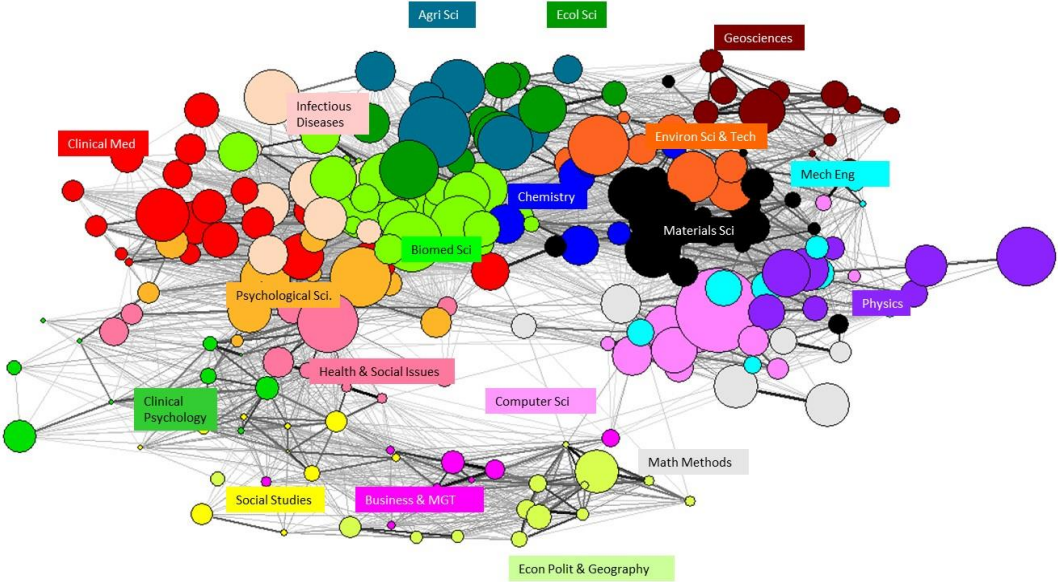

**Figure 3.** LAC map of Science, WoS Core Collection; 224 Web of Science Categories. Method based on Rafols, Porter and Leydesdorff (2010).

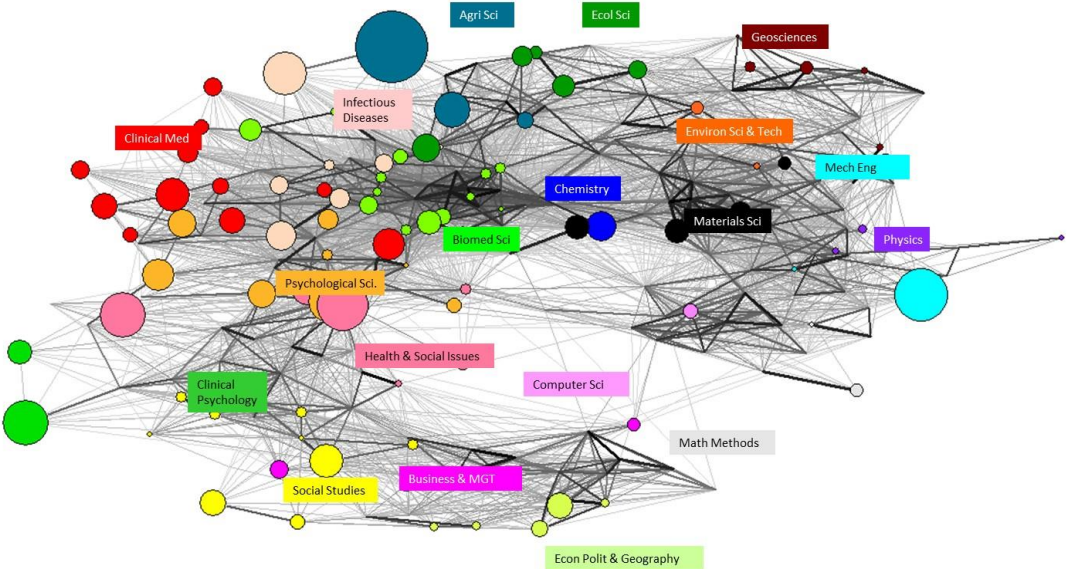

**Figure 4.** LAC map of Science, SciELO CI; 224 Web of Science Categories. Method based on Rafols, Porter and Leydesdorff (2010).

Note: According to Rafols, Porter & Leydesdorff (2010) method, the labels and colours in Figures 3 and 4 display 19 macro-disciplines (groupings of WCs) obtained using factor analysis of this same matrix. The size of nodes is proportional to number of publications.

Figures 3 and 4 make visible the differences in the thematic orientation of the communications in both indexes. Figure 3 shows major contributions in all categories from clinical medicine to physics and math methods which are better represented in the top of circular shape of WoS Core Collection. In Ffigure 4, we can observe that SciELO CI contains more journals in less categories: Social Studies in Yellow, Health and Social Issues and Clinical Medicine in pink and red respectively, Agricultural Science and Chemistry in aqua blue and blue respectively, Ecological Science in green, and Geosciences, Materials Sciences and Mechanical Engineering (brown, black, pale blue). In addition, WCs appear more disaggregated in Figure 4 than in Figure 3. The large number of journals contained in WoS explains the difference with SciELO CI in the visibility of SC.

The different disciplinary contexts from which SciELO and WoS originated might explain some of the differences between the regional and global scientific communications encountered in these two databases. Much has been written about the "natural" or hard-sciences origin of WoS; WoS originated from the Science Citation Index (Garfield, 1971), but has been expanded to include a broader range of journals and then enlarged by the Social Science Citation Index and later on by the Arts & Humanities Citation Index (the Science Citation Index was officially launched in 1964, the Social Science Citation Index followed in 1973, and the Arts and Humanities Citation Index in 1978). Meanwhile SciELO resulted from a cooperation which was formalized in 1997 between the Fundação de Amparo a Pesquisa do Estado do São Paulo (FASPEP) and the Latin American and Caribbean Center for Health

Sciences Information (Bireme) of the Panamerican and World Health Organizations
(PHO/WHO).

**Table 6. Volume of articles by WoS Categories in WoS and SciELO (Top 20 in SciELO)**

| SciELO | | | WoS | | | |
|---|---|---|---|---|---|---|
| Rank | $N$ | % | Rank WoS | $N$ | % | Id |
| 1 | 3017 | 10,8% | 52 | 1009 | 0,69% | Agriculture, Dairy & Animal Science |
| 2 | 1608 | 5,8% | 192 | 79 | 0,05% | Engineering, Aerospace |
| 3 | 1494 | 5,4% | 4 | 2284 | 1,56% | Public, Environmental & Occupational Health |
| 4 | 1186 | 4,3% | 85 | 633 | 0,43% | Education & Educational Research |
| 5 | 1141 | 4,1% | 100 | 521 | 0,36% | Nursing |
| 6 | 1036 | 3,7% | 14 | 1890 | 1,29% | Veterinary Sciences |
| 7 | 1026 | 3,7% | 145 | 246 | 0,17% | Psychology |
| 8 | 706 | 2,5% | 3 | 3063 | 2,10% | Plant Sciences |
| 9 | 698 | 2,5% | 165 | 171 | 0,12% | Sociology |
| 10 | 656 | 2,4% | 18 | 1731 | 1,19% | Surgery |
| 11 | 610 | 2,2% | 35 | 1291 | 0,88% | Dentistry, Oral Surgery & Medicine |
| 12 | 595 | 2,1% | 135 | 300 | 0,21% | Rehabilitation |
| 13 | 532 | 1,9% | 49 | 1031 | 0,71% | Chemistry, Analytical |
| 14 | 531 | 1,9% | 54 | 987 | 0,68% | Tropical Medicine |
| 15 | 502 | 1,8% | 74 | 695 | 0,48% | Health Care Sciences & Services |
| 16 | 488 | 1,8% | 10 | 2108 | 1,44% | Zoology |
| 17 | 478 | 1,7% | 78 | 684 | 0,47% | Sport Sciences |
| 18 | 470 | 1,7% | 43 | 1206 | 0,83% | Psychiatry |
| 19 | 422 | 1,5% | 139 | 296 | 0,20% | Anthropology |
| 20 | 408 | 1,5% | 117 | 416 | 0,28% | History |

Table 6 shows important differences between both databases which support the argument of differing thematic orientations. The documents in the data set were assigned to 99 subject categories in SciELO and to 204 (of the 250) WCs in WoS. There is also an important difference in the association of subject categories per journal: WoS journals have, on average, more subject categories assigned to them than SciELO CI indexed journals. The respective distributions show significant intellectual differences in the diffusion of regional versus global scientific knowledge produced in LAC, especially in the fields of Agriculture Sciences, Public Health, Social Sciences, and the Humanities. It is relevant to highlight that Aerospace Engineering has more presence in SciELO than in WOS, showing regional strengths in this field which are particularly clustered in Chile.

**Discussion and conclusions**

We used descriptive statistics about LAC contributions in journals indexed in WoS; our results suggest that SciELO CI integrates a scientific production which otherwise remains invisible in the mainstream journals contained in WoS. The perseverance in LAC scientific communications of Spanish and Portuguese as the main languages for communication, together with differences in the nature of the publishing venues, the geographical distribution of collaborations, and the disciplinary orientations of the contributions all seem to provide evidence suggesting that the integration of SciELO CI into WoS databases will allow a better representation of research capacities and strengths in LAC.

The collaboration networks analyzed, suggested that SciELO has in fact provided a platform for interactions among LAC researchers. As mentioned in the introduction, SciELO's open access policy relied on facilitating access to promote visibility. Open access, as a means to make visible research results that do not rise to the level of global interests but that might be relevant to countries with similar problems, has been part of the policy agenda for a while (e.g., Wagner & Wong, 2011). Contributions in SciELO-CI indexed journals have reached beyond the LAC region to include authorships from Africa and Asia, suggesting an interesting data set to study South-South collaboration.

Collaborations in LAC contributions included in WoS-indexed journals are more frequently mediated by the more developed countries' capacities, particularly from Europe and the USA. Nevertheless, researchers from LAC countries have a primary role as first authors in 2/3 of the multi-authored papers. This means that LAC researchers are well embedded in the global scientific dynamics.

The distribution of contributions in terms of WoS Subject Categories show that SciELO CI differs in its coverage of disciplines and specialties from WoS. This was illustrated (in Figures 3 and 4) using overlays of the two datasets with LAC authors on the same basemap. SciELO CI seems then to be better at representing scientific contributions where the particularities of the region and the social context are important. An exercise exploring aggregated journal-journal citation relations in the Chinese Science Citation Index of the Library of the Chinese Academy of Sciences found that the high frequency of university-based journals in the index provided a practical ends-based structure more aligned to Mode 2 knowledge production (Leydesdorff & Bihui, 2005). Although such a study using SciELO CI would be difficult due to the lack of journal-journal citation information at this point, the frequency of academic publishing sources in SciELO CI indexed journals might provide a similar intellectual organization to the regional journal structure.

The inclusion of SciELO CI into WoS responds to the need for a more inclusive representation of scientific results despite regional constraints and conditions. It may also reflect increased competition for the services offered by Thomson Reuters and Elsevier. However, the strategies aimed at improving regional visibility seem to differ between Scopus and WoS. While Scopus has aimed at increasing their base of regional journals, the globalization of the Web of Science (Testa, 2011) has also meant the incorporation of regional databases as a whole and not on the basis of evaluating individual journals. The Chinese Journal Database has been hosted in the WoS since 2008, while the inclusion of SciELO CI and the Korean Journal Database has been operative since 2014.

From a technical point of view, this inclusion opens the door to a new research agenda. Before the integration of SciELO CI into WoS, the alternatives to using SciELO CI-data for bibliometric studies were limited. Although SciELO's program relied on the importance of Open Access to increase the visibility of scientific results, the platform did not provide appropriate tools to download data, nor did it allow for the simple analysis of results as provided by WoS. These new opportunities for bibliometricians will also improve some challenges for the editors of SciELO CI-indexed journals. The inclusion of SciELO CI into WoS should, in the short to mid-term, improve to compliance with international editing norms and governance structures. Editors of international journals position their journals by generating the quality, both editorial and cognitive, of the contents of their journals. Competition for relevant contents as well as a better evaluation of the referencing procedures will probably be increasingly important for the agendas of LAC journals. We would like to explore this issue further in order to understand how the inclusion of SciELO CI might restore the WoS to the competition for visibility of regional results, as well as improving the quality of the LAC journals included in SciELO CI.